\documentclass[sigconf, nonacm]{acmart}

\usepackage[utf8]{inputenc}
\usepackage[english]{babel}
\usepackage{lmodern}
\usepackage[T1]{fontenc}
\usepackage{algorithmic}
\usepackage{graphicx}
  \graphicspath{{images/}}
\usepackage{subcaption} 
\usepackage{tabularx}
\usepackage{textcomp}
\usepackage{xcolor}
    
 \AtBeginDocument{%
  \providecommand\BibTeX{{%
    \normalfont B\kern-0.5em{\scshape i\kern-0.25em b}\kern-0.8em\TeX}}}

\newcommand{\eg}{e.\,g.\ }

\begin{document}

\title{A Case for Practical Configuration Management Using Hardware-based Security Tokens}

\author{Tim Lackorzynski}
\affiliation{%
  \institution{TU Dresden}
  \city{Dresden}
  \country{Germany}
}
\email{tim.lackorzynski@tu-dresden.de}

\author{Max Ostermann}
\affiliation{%
  \institution{TU Dresden}
  \city{Dresden}
  \country{Germany}
}
\email{max.ostermann@mailbox.tu-dresden.de}

\author{Stefan K\"{o}psell}
\affiliation{%
  \institution{Barkhausen Institute / CeTI TU Dresden}
  \city{Dresden}
  \country{Germany}
}
\email{stefan.koepsell@barkhauseninstitut.org}

\author{Hermann H\"{a}rtig}
\affiliation{%
  \institution{TU Dresden}
  \city{Dresden}
  \country{Germany}
}
\email{hermann.haertig@tu-dresden.de}

\renewcommand{\shortauthors}{Lackorzynski et al.}

\begin{abstract}
Future industrial networks will consist of a complex mixture of new and legacy components, while new use cases and applications envisioned by Industry 4.0 will demand increased flexibility and dynamics from these networks.
Industrial security gateways will become an important building block to tackle new security requirements demanded by these changes.
Their introduction will further increase the already high complexity of these networks, demanding more efforts in properly and securely configuring them.
Yet, past research showed, that most operators of industrial networks are already today unable to configure industrial networks in a secure fashion.

Therefore, we propose a scheme that allows factory operators to configure security gateways in an easy and practical  way that is also understandable for staff not trained in the security domain.
We employ hardware security tokens that allow to reduce every day configuration to one physical interaction.
Our results show the practical feasibility of our proposed scheme and that it does not reduce the security level of industrial security gateways in any way.
\end{abstract}

\keywords{Industry 4.0, Industrial IoT, IIoT, Industrial Automation, Industrial Networks, Security, Usability, Middlebox Security}

\maketitle

\section{Introduction}

Industry 4.0 will change the landscape of factory networks.
Various new use cases will require new technologies resulting in more heterogeneity.
The amount of networking will increase within the factory as well as with cloud services outside of the factory.
Additionally, the long life cycles of industrial machinery will lead to a side-by-side of new and legacy equipment \cite{nist_guide_to_ics_security}.
Factory networks will consist of devices and smart services serving different use cases being provided by many different vendors.
The security of these networks will become a top priority, as these systems get connected to the Internet and therefore reachable to malicious actors. Security incidents of the past may serve as warning signs of this general trend. The NotPetya attack may serve as a powerful example \cite{notpetya}.

\begin{figure}[t]
\centering
\includegraphics[width=\columnwidth]{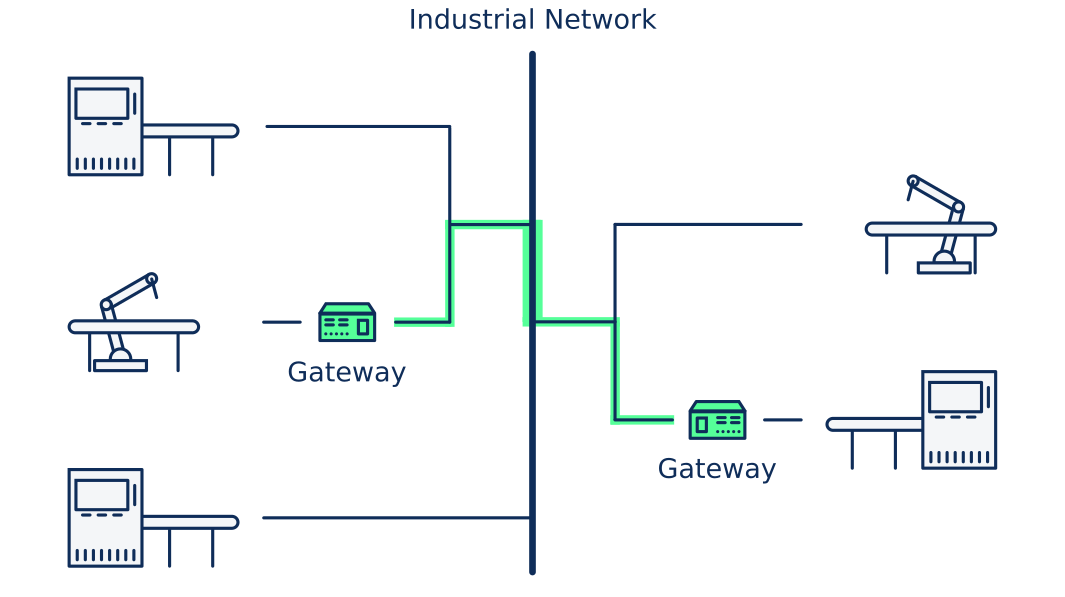}
 \caption{An industrial network with a heterogeneous set of devices. The green encryption gateways protect machine traffic by establishing a secure channel.}
 \label{fig:network}
\end{figure}

One approach to improve the security of these heterogeneous networks is deploying encryption gateways or middleboxes in the network \cite{fastvpn, remote_vpn_to_ics, poor_mans_fastvpn}. 
These are put between an industrial machine and the network, as shown in Fig.~\ref{fig:network}.
Secure channels can then be configured between the gateways so that all network traffic between the respective machines is encrypted and protected from the rest of the potentially insecure or even malicious network.

However, these gateways then also represent an additional infrastructure that adds another layer of complexity, which has to be maintained and managed.
Security management is typically rather complex and therefore error-prone.
Configuration errors even happen to expert staff leading to vulnerable networks \cite{bad_config_opc_ua}.
Additionally, security measures are mostly seen by non experts as overheads, typically leading to setups, where security measures are configured in such a way that they interrupt the productive work the least. This often leads to configuration rules that do not restrict behavior in any way and hence offer no practical increase in security.
This is especially true in the industrial environment, where staff is classically often not trained in IT security.
As factory networks will become more dynamic through Industry 4.0-related shifts in technology, these problems will only increase in the future.
For example, the vision of mass customization down to a lot size of one will demand highly flexible networks, where reconfiguration will be frequent.
Any future security scheme must consider this.
Consequently, understandability and usability will be key factors to the successful implementation of additional security measures in general and encryption gateways in particular.

This paper presents a novel approach on how to securely configure said encryption boxes.
We aim for a solution that is very understandable and actionable by staff that is not trained in IT security matters.
To this end, we use hardware-based security tokens, that reduce the configuration of a secure channel to one physical action that does not require further interaction with any software user interface.

The remainder of this work is structured as follows.
Sec.~\ref{sec:background} gives more insights into the importance of usability for security related mechanisms in industrial environment and details our general approach.
Sec.~\ref{sec:related_work} surveys some related work.
Sec.~\ref{sec:design} first motivates our design by presenting some goals, we based our design on, and then presents our scheme.
A prototypical implementation of our scheme is presented in Sec.~\ref{sec:implementation}.
Sec.~\ref{sec:evaluation} evaluates the design as well as the implementation.
Sec.~\ref{sec:conclusion} concludes the paper.
  
\section{Background}
\label{sec:background}

Today, many industrial systems and factory networks are vulnerable due to bad configuration.
A recent study by Dahlmanns et al. showed that Internet-facing industrial communication systems are frequently configured in a way, that makes them or the end points that they connect to susceptible to attacks \cite{bad_config_opc_ua}.

In more detail, the researchers scanned the whole Internet (whole IPv4 address range) for OPC UA servers. OPC UA is a modern service-oriented communication protocol for industrial machines that among other things also offers state of the art encryption \cite{opcua}.
The authors showed that 26\% of the OPC UA servers connected to the Internet were configured without any access restriction. Further 25\% of servers used a deprecated and now considered insecure hash function (SHA-1).
On many servers that used certificate-based authentication, the study found that identical certificates were used, probably due to the operators of these systems just copying the certificate from another server during provisioning. 
The manual provided by the vendor of the OPC UA server was obviously not consulted.
The authors of the study then went on to alarm the vendor, who in turn notified his customers, the operators of the servers, about this security incident. Yet, this did not result in an update of those systems.
Additionally, 44\% of all OPC UA servers on the Internet allowed unauthorized users to read and write values from industrial devices and execute system functionality.
Only 8\% of systems were configured correctly.
In summary, this study shows a deep divide between the potentially achievable security levels offered by modern industrial communication systems and the actual level of security in deployed systems in practice by virtue of bad configuration of the available security mechanisms.
Additionally, the study hints at a lack of understanding about network security topics at the side of the factory operators as they did not update their vulnerable systems even after being informed about the weakness.
As this study scanned the whole Internet, these problems must be considered industry-wide. 

\begin{figure}[t]
\centering
\includegraphics[width=\columnwidth]{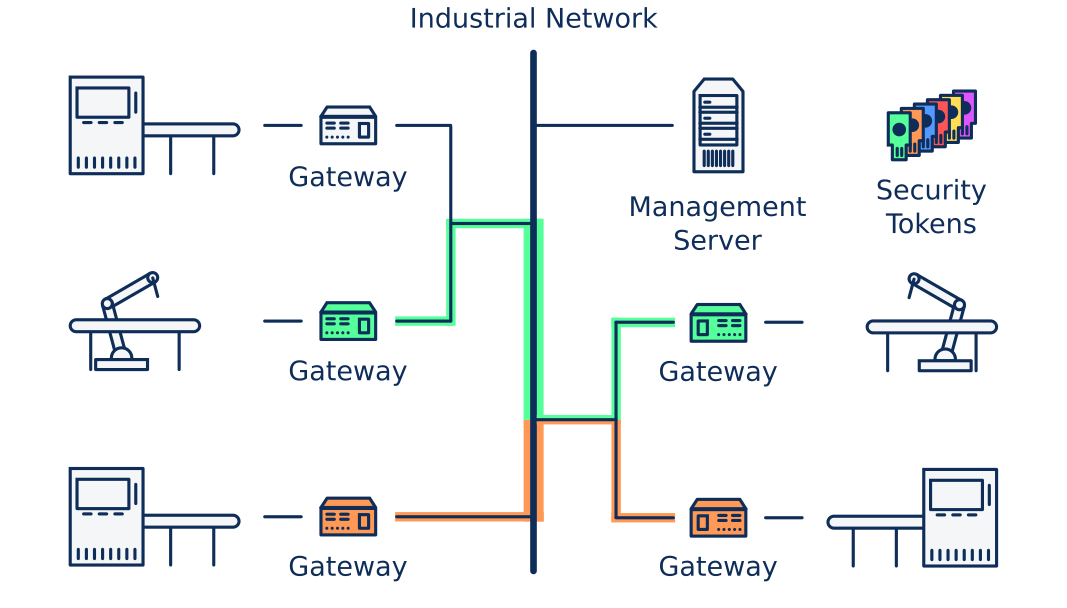}
 \caption{An industrial network where different secure channels protect communication traffic of end points. The gateways are managed by a central server. Secure channels can be configured via security tokens.}
 \label{fig:secured_network}
\end{figure}

As already described above, encryption gateways are computing nodes that are added to the network and provide an additional layer of security.
To be effective, numerous gateways have to be distributed within the factory, which then require some kind of mechanism that allows for regular reconfiguration of said secure channels.
For our scenario, we assume a central server, that manages the deployed gateways, see Fig.~\ref{fig:secured_network} for reference.
Secure channels can be configured dynamically depending on the current work load of the factory and there can be multiple channels active at the same time. Although not pictured, a single secure channel can be configured between more than two gateways. A gateway, where no secure channel is configured, does not touch the traffic and just transparently patches it through.

The standard way to configure such as setup would be a management software with some graphical user interface (GUI) running on the server, where gateways are listed and an operator can configure the secure channels.
In a best case scenario, the gateways are represented with some form of graphical icons, but more typically these interfaces are rather text-heavy and based around drop-down menus. 
In any case, this type of interface is already pretty abstract, virtual and divorced from the actual physical end points that shall be protected.
Normal functional configuration of factory equipment is already complicated and any security measure, like encryption gateways, add yet another layer of complexity to these networks.
Additionally, the secure channels are going to be frequently reconfigured as these are dependent on the concrete work load or application the factory is executing at a certain time.
Being able to configure them with ease is therefore paramount for the acceptance and adoption of encryption gateways in industrial networks as a whole.

The easiest and most comprehensible way to configure these secure channels would be by using a physical token, which is plugged into each respective gateway.
This would mimic established security protocols from the real world, namely locking a door with a key, where the key is the secret, that needs to be protected. This would immediately make sense for anyone, including people not classically trained in IT security matters.
We believe, this has the potential to increase the actual security level of factory networks more closely to what is theoretically possible.
Therefore, we propose a configuration solution for encryption gateways using physical keys or tokens to configure the secure channels between them.
This scheme is not supposed to supplant classic GUI-based configuration schemes, but to augment and improve them. 
Consequently, we still assume a graphical configuration interface on the central management server.

\section{Related Work}
\label{sec:related_work}

Hardware security modules (HSMs) are a special class of devices.
They are fully independent computing devices equipped with their own microprocessor and memory.
They can store secrets safely and use them for cryptographic functions to provide various security functionalities, like issuing digital keys, encryption and decryption of data, digital signatures, secure storage of data or authentication. 

A specific subgroup are security tokens. 
These are used to authenticate its holder to gain access to some type of resource.
Therefore, they are also called electronic keys.
They range from wireless key cards for opening doors to tokens provided by banks offering additional protection for online banking services.

A common use case is to bind a software product to a so-called security dongle. These are hardware tokens that have a certificate stored and which have the ability to attest the existance of that certificate without it having to leave the token. These dongles must be plugged into the device wishing to execute the bound software application.
Entire business solutions are readily available, that comprise of dongles, software libraries and cloud authentication services. The CodeMeter product line from the company Wibu-Systems might serve as good example \cite{wibu_codemeter}.

Yet, these types of approaches have in common that they only support a static mapping of token to resource.
Tokens are just used as physical embodiments of a digital certificate and their presence always unlocks the same resource.
The tokens of the scheme presented in this work on the other hand, must be able to be dynamically mapped to different resources (secure channels) and interact with multiple devices (encryption gateways) interchangeably.

Another common use case for security tokens is two-factor authentication (2FA).
There, users are only granted access to a remote service, when they provide a password as well as a second type of identification. In this case, a token that can authenticate itself against the remote server. The application of 2FA in online authentication processes has been increasing in the last years, as more and more online service providers start offering this type of remote authentication.
Yet, to the knowledge of the authors, there is no work that employs those security tokens in a similar usability concept as proposed here, not in the industrial environment nor anywhere else.

The closest work we could find, was a commercial industrial router, where a classic physical key has to be plugged in and turned to a certain position to allow a remote party to access the network \cite{mbnet_rokey}.
The physical action of turning the key is used to signify the unlocking of the network behind the router.
This goes in the same direction as our approach in that it reduces a complex security configuration to a simple physical action.
Yet, since this is just a normal physical key, it can easily be copied and, should it get stolen, the router has to be replaced.
Furthermore, this approach does not cover our use case, as we want to propose a solution for encryption gateways and not routers.

\section{Design}
\label{sec:design}

Section~\ref{sec:goals} will present certain design goals derived from the observations above. These goals on the one hand try to formulate in detail why and where exactly security measures must be applied to produce trust in the overall system. On the other hand they try to formulate what exactly is meant with the general goal of usability.
These design goals then will form the cornerstones for our design, which will be presented in Sec.~\ref{sec:scheme} below.

\subsection{Design Goals}
\label{sec:goals}

To achieve the goal of being able to configure secure channels on the encryption gateways using only physical tokens, certain preconditions must be met, that revolve around establishing trust relationships between the distributed entities of the system (server, gateways, tokens).
Therefore the following design goals include the establishment of trust, based on which further design goals can be defined.

\subsubsection{Trust between Server and Gateways}

The encryption gateways are scattered within the factory and are connected via the factory network to the server.
We consider this network insecure, otherwise there would be no need for encryption gateways in the first place.
Therefore, the gateways must build a trust relationship with the server, so that they can authenticate themselves when connecting to the server remotely.
This trust can then form the basis for a secure management channel, where the server receives status updates from and can send configurations to the gateways.
  
\subsubsection{Trustworthy Tokens}

The main idea of our scheme is that just a token should suffice to establish (or tear down) a secure channel between two gateways.
Therefore, these tokens must not be forgeable and it should not be possible to just copy or replicate them.
To this end, they should have some kind of unique identifier so that they can always be also physically accounted for.
Furthermore, they should be able to authenticate themselves remotely against the management server so as to be able to build a trust relationship.
  
\subsubsection{Trustworthy Configuration}

When trust between the server and the gateways as well as between the server and the tokens has been established, the scheme must make sure, that only trustworthy tokens connected to trustworthy gateways are allowed to change the configurations of secure channels.

\subsubsection{Ease of Use}

The everyday use, meaning the setup and tear down of secure channels should be as easy as possible.
One physical action should be enough and no interaction via some software interface should be necessary.

\subsubsection{Life-cycle Management}

Any scheme of that kind can only hope to be implemented in practice, if it is possible to manage the whole life-cycle of all the components.
This means that it should be possible to also remove gateways and tokens from the system without compromising the security of other still connected parts.
Especially, the scheme should be able to deal with a token getting lost, broken or potentially stolen.

\subsubsection{Attacker Model}

The general aim of our scheme is to increase the usability of encryption gateway-based systems.
Yet, security still has the highest priority and the security of the encryption gateways must not be compromised in any way.
Therefore, no new attack vectors shall be opened and increases in attack surface shall be as minimal as possible.

For our design, we assume an attacker that has physical access to the network.
He can steal tokens and maliciously use them inside the factory.
Yet, he cannot copy them or forge fake tokens.
We do not assume the attacker to be able to physically remove the encryption gateways or manipulate the network infrastructure.
If we granted the attacker this power, any security scheme would be moot, as he could simply remove all gateways or mount a man-in-the-middle attack between the encryption gateway and the individual end point the gateway tries to protect.
The protection of the physical infrastructure is a concern of operational security and outside the scope of such a scheme as described here.
For example, the removal of a life encryption gateway could be detected by a missing heartbeat signal.
A subsequent automatic alarm issued by the system could then alert personell to handle the situation.

\subsection{Scheme}
\label{sec:scheme}

Our scheme consists of multiple steps that establish trust between the individual components involved in the system. This binds them together and as a result easy configuration of secure channels between gateways becomes possible.
All steps are presented in the following, while some are also depicted in Fig.~\ref{fig:steps} for additional clarity:

\begin{figure}[t]
\centering
\begin{subfigure}{0.49\columnwidth}
\centering
\includegraphics[width=1.0\columnwidth]{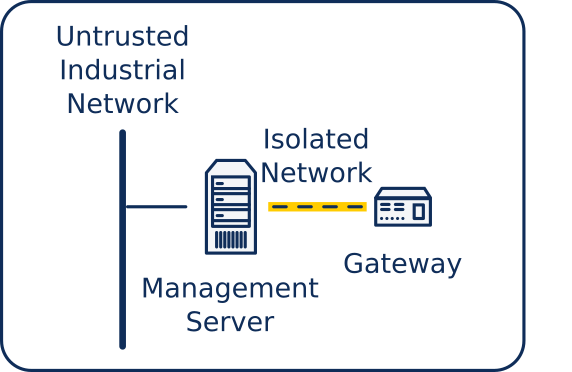}
 \caption{Gateway provisioned by server.}
 \label{fig:step1}
\end{subfigure}
\begin{subfigure}{0.49\columnwidth}
\centering
\includegraphics[width=1.0\columnwidth]{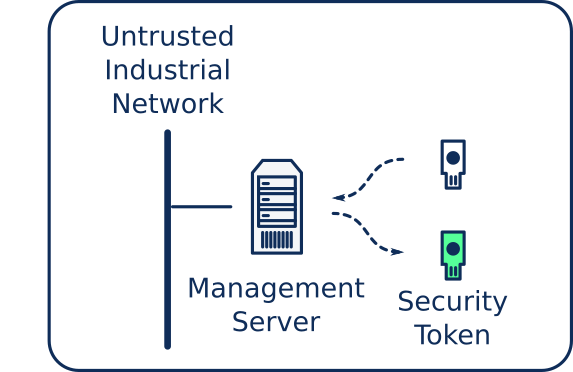}
 \caption{Token provisioned and bound to green secure channel.}
 \label{fig:step2}
\end{subfigure}
\\
\begin{subfigure}{1.00\columnwidth}
\centering
\includegraphics[width=1.00\columnwidth]{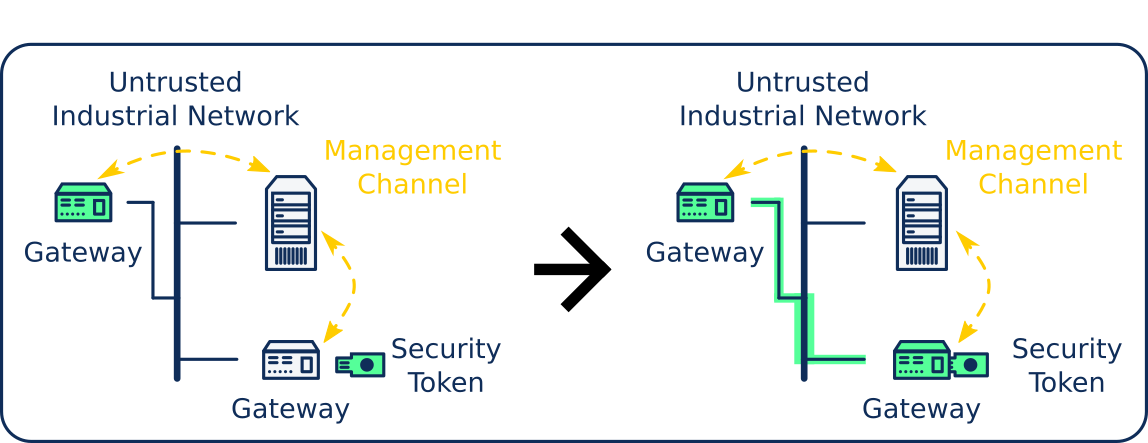}
 \caption{Configuration of a secure channel on a deployed gateway via security token.}
 \label{fig:step3}
\end{subfigure}
 \caption{Steps 1 to 3 of our scheme.}
 \label{fig:steps}
\end{figure}

\paragraph{1. Establishing a Trust Relationship between Server and Gateway}
\label{sec:step1}

Before a gateway can be deployed in the network, there must be an initial phase, where cryptographic keys (i.~e. public keys, pre-shared keys or certificates) are exchanged between the gateway and the server, so that later mutual authentication and the establishment of a secure management channel over the network becomes possible. 
This exchange necessarily happens without any security as there is no secure channel established yet.
Therefore, during this provisioning phase, the gateway must be connected to the server in an isolated environment, where it can be guaranteed that there is no attacker present.
This can be realized by connecting the gateway to a dedicated physical port on the server (\eg USB or Ethernet) as shown in Fig.~\ref{fig:step1}, or by putting both in a separate isolated network.
In more complex settings, an offline backup server can be used to onboard gateways. The exchanged credentials are then mirrored back to the live server via a single direct channel.
Irrespective of how it is concretely done, the gateway can then be deployed in the network and can establish a secure management channel with the server over the network.
  
\paragraph{2. Establishing a Trust Relationship between Server and Token}
\label{sec:step2}

The token must also go through a provisioning phase, where it is directly connected to the server so that secrets can be securely exchanged, making later remote authentication possible.
Additionally, the token's identifier must be registered by the server and the server must bind this token identifier to some secure channel identifier as shown in Fig.~\ref{fig:step2}. This also makes it possible to bind multiple tokens to the same secure channel.
  
\paragraph{3. Setup of a Secure Channel on a Gateway}
\label{sec:step3}

Once trust relationships are established and the gateways are deployed, the actual secure channels (green in Fig.~\ref{fig:step3}), that are supposed to protect the live communication traffic of the end points, can be configured.

First, a token must be plugged into the gateway in question.
After a physical user interaction, like pressing a button on the token, the token emits its token identifier and some secret necessary for authentication against the server.
Both are then sent via the secure management channel to the server.
The server can then validate all information. It can prove that the token is trustworthy and known and that it is connected to a trustworthy gateway.
It then registers the gateway as a new participant in the secure channel bound to the token identifier.
After that, it sends back all necessary information for the gateway it needs to configure the secure channel.
This might include network addresses of other participants and security keys or certificates.
It also updates other participants of the secure channel about this new participant.
All gateways proceed to update their network interface configurations accordingly.
Finally, the newly configured gateway indicates its new state by some means to give feedback to the operator.
This might be implemented by for example signal LEDs or an attached display.
The gateways are now able to exchange encrypted traffic from their respective endpoints, as depicted in Fig.~\ref{fig:secured_network}.

\paragraph{4. Tear Down of a Secure Channel on a Gateway}
\label{sec:step4}

To tear down the secure channel on the gateway, the gateway must be removed from the list of participants on the server.
This can be achieved in two ways.
First, the gateway can be removed by using the corresponding token again.
It is plugged in and everything happens exactly like described in the previous step, only that now the gateway is deregistered from the secure channel. All remaining participants are updated and they configure their network interfaces accordingly.
The gateway removes the secure channel from its network configuration and goes back to its default behavior, which for example, might be to transparently patch through traffic.
The second way to remove a gateway is to just issue a command to deregister it directly on the server. Everything else then happens just as described above.
  
\paragraph{5. Decommissioning of a Gateway}
\label{sec:step5}

Should a gateway need be removed or replaced because it broke or was successfully attacked, the server can just order other participants to stop interacting with this gateway by deregistering it from all its secure channels.
Other gateways will then simply ignore this one.
  
\paragraph{6. Decommissioning of a Token}
\label{sec:step6}

In the same manner as in the previous step above, a token can be decommissioned by removing the binding of its token identifier to the secure channel identifier.
If the token was lost or is feared to have been stolen, it would also be necessary to tear down the associated secure channel on all participants, as described in Sec.\ref{sec:step4}. A new channel replacing the old could then be configured using a new token.

\section{Implementation}
\label{sec:implementation}

\begin{table*}
\begin{center}
\begin{tabularx}{\textwidth}{|c|p{3.0cm}|X|p{6cm}|}
\hline
\textbf{Step} & \textbf{Token (T)} & \textbf{Management Server (MS)} & \textbf{Gateways (GW)}\\
\hline
1. & --- & GW = (pubKey\textsubscript{GW}, IP Address\textsubscript{GW}, MS~Address\textsubscript{GW}), privKey\textsubscript{MS} & pubKey\textsubscript{MS}, IP Address\textsubscript{MS}, privKey\textsubscript{GW}\\
\hline
2. & tokenID, OTP Secret & T = (tokenID, OTP Secret), & ---\\
 & & Secure Channel = (secID, MS key, \{T\}, *) &\\ 
\hline
3. & \textit{(tokenID, OTP)} & Secure Channel = (secID, MS key, \{T\}, \{GW\}) & secID, \{MS Address\textsubscript{GW}\}, MS key\textsubscript{secID}\\
\hline
\end{tabularx}
\end{center}
\caption{Exchanged information between entities per step. Data in cursive is ephemeral and only used once.}
\label{tab:hw_info}
\end{table*}

This section discusses our prototypical implementation, and follows the same step structure that was introduced above.
Tab.~\ref{tab:hw_info} shows which entity in our scheme holds which information and in which step that data is exchanged.

We implemented the server and the gateways using Raspberry Pi 4 minicomputers running Ubuntu Linux 20.04.

The trust relationship between server and gateways as described in Sec.~\ref{sec:step1} was realized using Wireguard \cite{wireguard}.
It is a modern, secure and easy to use IP layer encryption protocol that is superior to older comparable solutions \cite{my_vpn_study}.
Using Wireguard provides us with a state-of-the-art protocol that is agnostic towards the transport layer and the concrete implementation of the management protocol.

The secure channels being configured in Step 3 were implemented using MACsec \cite{macsec}.
This encryption scheme allows to protect whole Ethernet frames, which makes it agnostic or transparent to the upper layer protocols. This makes it a good choice for the heterogeneous field of industrial communication \cite{my_macsecpp}. Additionally, it fits our approach, as it can be configured on a peer-to-peer basis.

\paragraph{1. Establishing a Trust Relationship between Server and Gateway}

We implemented the isolated channel by connecting the gateway directly to a dedicated network port of the server.

To setup the management channel using Wireguard, both communication partners each need to generate a public/private key pair. The public keys (pubKey) are exchanged, while the private keys (privKey) are stored on the device.
Additionally, both partners exchange IP addresses, enabling them to establish the management connection later, when the gateway is deployed in the network.

The management server must also store MACsec addresses (MS address) for each gateway.
These are different from the IP addresses of the management channel and are necessary for the MACsec software clients running on the gateways to be able to connect to each other. The addresses are disseminated in a later step to gateways joining a secure channel.

We implemented our own rudimentary management protocol using gRPC\footnote{https://grpc.io/}.
All configuration steps detailed here were automated using scripts written in Python or bash if not otherwise stated.

\paragraph{2. Establishing a Trust Relationship between Server and Token}

\begin{figure}[t]
\centering
\includegraphics[width=0.3\columnwidth]{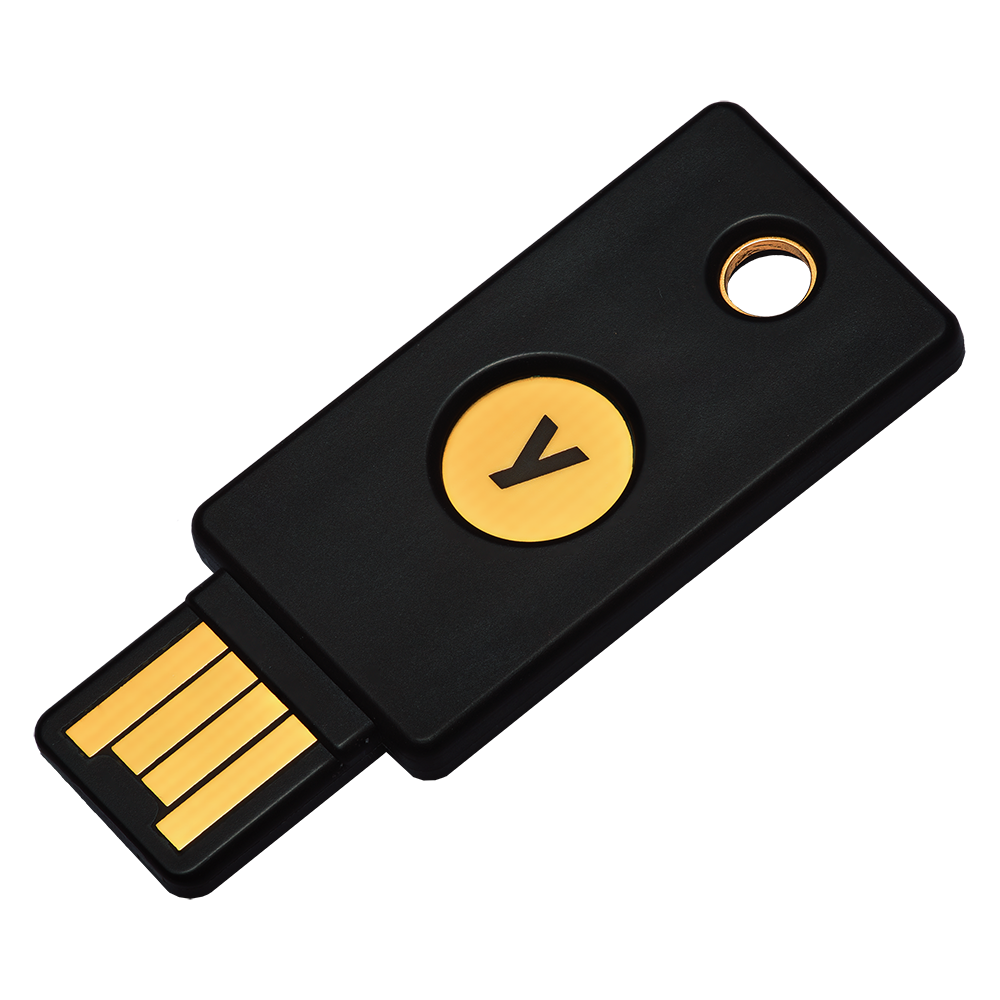}
 \caption{A YubiKey FIPS security token, as was used in our prototypical implementation.}
 \label{fig:yubikey}
\end{figure}

\begin{figure}[t]
\centering
\includegraphics[width=0.2\columnwidth]{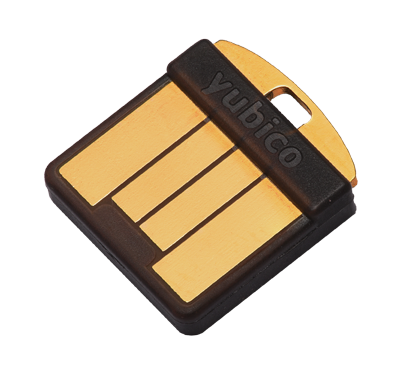}
 \caption{A YubiHSM 2 hardware security module, as was used for secure storage of secrets on the server.}
 \label{fig:yubihsm}
\end{figure}

We implemented the hardware security tokens using the YubiKey FIPS\footnote{https://www.yubico.com/de/product/yubikey-fips/}, depicted in Fig.~\ref{fig:yubikey}.
These tokens can produce one-time passwords (OTPs), which are basically secret numbers that can be used once to authenticate a transaction against a remote station (the management server in our case). The YubiKey is plugged into a USB port and issues one OTP when the button at the center is pressed.

For the remote authentication to work and to establish the trust relationship described in Sec.~\ref{sec:step2}, corresponding OTP secret keys (OTP secret) from which individual OTPs are derived, must be provisioned on both the token and the remote station.
We used a software tool provided by the vendor for this purpose\footnote{https://www.yubico.com/products/services-software/download/yubikey-personalization-tools/}.

Furthermore, we use a YubiHSM 2 module\footnote{https://www.yubico.com/de/product/yubihsm-2/}, depicted in Fig.~\ref{fig:yubihsm}, attached to the management server as secure storage for the OTP secret keys.
Received OTPs can be checked against stored secret keys via an API. As a result, the security-critical secret keys are only stored on the employed hardware security modules and never on the server or gateways.

Additionally, each YubiKey comes with a unique serial number, which we use as the token identifier (tokenID).
Furthermore, the token is bound to a certain secure channel (\eg to green as depicted in Fig.~\ref{fig:step2}), which will be the one configured on the gateways later. Each secure channel is denoted by a secure channel identifier (secID).

Additionally, each time a new secure channel is configured, a new MACsec encryption key for that channel is also created (MS key).
For our prototypical implementation, we chose to configure MACsec using simple symmetric pre-shared keys.
  
\paragraph{3. Setup of a Secure Channel on a Gateway}
\label{sec:step3i}

Establishing a secure channel on the gateway works by inserting a provisioned YubiKey into a USB port of the gateway and pressing the button.
The YubiKey issues one OTP, which is then sent by the gateway together with the token identifier to the server via the management channel.

The server replies with all necessary configuration information for the secure channel bound to the YubiKey's token identifier.
This includes the secure channel identifier (\eg ``green''), MACsec addresses of all participants in that secure channel as well as the shared encryption key.
The server then updates all other participants of this secure channel with the necessary information about the appearance of another communication partner.

Finally, the gateway configures a corresponding MACsec interface on itself and sets up the necessary virtual network bridges.

\paragraph{4. Tear Down of a Secure Channel on a Gateway}
\label{sec:step4i}

Removing gateways from a secure channel works the same way as described in the previous step, only that the server removes the gateway from the secure channel (and not adds it) and updates the other participants to that effect. The gateway then modifies its network configuration accordingly.

The scheme also allows for a direct removal of a gateway via the local configuration interface of the management server.

\paragraph{5. Decommissioning of a Gateway}

The decommissioning of a gateway was implemented directly on the server by triggering the removal of said gateway from all configured secure channels.

\paragraph{6. Decommissioning of a Token}

The decommissioning of a token was implemented directly on the server by removing the binding of the token identifier to any secure channel.

\section{Evaluation}
\label{sec:evaluation}

In the following, the design of our scheme, presented in Sec.~\ref{sec:scheme}, as well as the implementation, detailed in Sec.~\ref{sec:implementation}, are compared to the design goals presented in Sec.~\ref{sec:goals}.
The first design goal was to have a trustworthy relationship between the management server and the gateways, so that the gateways could be centrally managed.
We used Wireguard to set up and protect the management channel and  as long as the server only reacts to management and configuration calls from the gateways from this local Wireguard interface, potential attackers cannot even address the service managing the gateways.

The second design goal was to be able to trust the hardware token, that are used for the configuration of the secure channels. The employed YubiKeys offer unique identifier and can be used to establish an OTP mechanism with the management server, effectively providing protection against being forged, copied or impersonated.

The third design goal was a trustworthy configuration.
Since the server only reacts to calls from the management interface that carry a proper OTP, only trustworthy modifications to the configuration are done.

The fourth design goal was to provide a solution, where the everyday use was as simple as possible.
While the bootstrap of the scheme is still fairly complex, the configuration of the actual secure channels is very easy and can be accomplished by just physically plugging a token into the gateway and pressing the button once.
The complexity of the bootstrapping process compares to other approaches, as establishing trust in distributed systems just demands a certain complexity that cannot be optimized any further. Secrets have to be exchanged and network configurations have to be set. Yet, this has to be done only once for each deployment of a gateway or token, which should be fairly rarely. This can be delegated to a domain expert.
The configuration of secure channels on the other hand can be accomplished by even in IT security matters untrained staff.

The fifth design goal stated that the solution should allow for life-cycle management.
Our design accounted for this, by including steps in the design, where gateways and tokens could be decommissioned, even without their cooperation.
Broken or compromised components cannot compromise the system as their information can be removed from the databases directly.
Renewal of components is possible by decommissioning them and adding new ones. In case of a token, a new token can even be bound to the same secure channel, the previous token was bound to, so that it is not necessary to interchange the secure channel, when a token is replaced.

The last design goal stated, that while usability was the aim of this scheme, the security of the encryption gateway-based system to which our token-based scheme is only attached to, still has paramount priority.

From the software prospective, the security of our scheme hinges on the security of the management server, but that is already the case, even without our hardware-based security scheme on top. We just added some more functionality and responsibility, but as the server already manages encryption gateways, it is already considered to be a high value target and should therefore be engineered and deployed with the utmost consideration for operational security.

Modern management servers as well as the encryption gateways are in practise equipped with general purpose operating systems that support the application layer with rich functionality.
The overall amount of software employed in this scenario is huge and our scheme only adds very small amounts of complexity in comparison.
A fully realized software stack, that is necessary to implement such a use case, consists of 10s to 100s of millions of lines of code (LoC), while the scripts necessary to implement our scheme, and which are additionally executed by the management server and the gateways, only comprise of \textasciitilde600 LoC.
From the software perspective, the increase of attack surface by our scheme is negligible.

Yet, the scheme also introduces additional hardware tokens that can be apprehended by an attacker inside the factory. And while he can use them to change or delete gateways from secure channels to disturb the functioning of the factory, he cannot divert protected traffic outside the factory, as tokens alone are not sufficient to establish a secure connection. He would need a properly configured gateway as well and we assume he cannot easily steal one.
Also, while the attacker can maliciously configure secure channels using a stolen token, this configuration would be registered via the regular mechanisms of our scheme. A factory operator then has the chance to rectify this action by canceling the issued command and by removing the stolen token from the system by means of its token identifier, as it is not necessary for the operator to be in possession of the physical item to do so.
The token identifier is either found in the system as the issuer of the malicious command or in some (paper) file that was filled in when the token was physically issued to a factory worker, which reported its loss.
From this prospective, our scheme also increases the attack surface.
Yet, it is again minimal, as malicious actions can be monitored in real-time and be reverted instantly. An attacker can hence only incur temporary harm.
How long this temporary phase is, depends on operational considerations within the factory and is outside of our scope.

\section{Conclusion}
\label{sec:conclusion}

In the light of increasing complexities in future smart factories, the actual security of factory networks will strongly depend on the proper deployment of specific protection mechanisms.
This deployment hinges on two factors. First, factory operators, who are no domain experts for IT security, must understand the protection mechanism and the mechanism itself must be as frictionless as possible so as to not impede the productive work of the factory.

In this work, we made the case for a novel mechanism that allows to easily and understandably configure encryption gateways, which will be important building blocks for the security architecture of future factories.
We implemented a configuration scheme that employed hardware security tokens. These tokens cannot be forged and reduce the actions necessary to implement a security policy to a simple physical action, comparable to locking a door with a key.
Our results showed, that a reduction of the complexity of configuration must not necessarily be accompanied with a loss of security.

\begin{acks}
This work was co-funded by the SAB (Development Bank of Saxony) under frameworks from both the ERDF (European Regional Development Fund) as well as the ESF (European Social Fund), by public funding of the state of Saxony/Germany and by the German Research Foundation (DFG) as part of Germany’s Excellence Strategy EXC 2050/1 – Project ID 390696704 – Cluster of Excellence Centre for Tactile Internet with Human-in-the-Loop (CeTI) of TU Dresden.
\end{acks}

\bibliographystyle{ACM-Reference-Format}

\end{document}